\documentclass[12pt]{iopart}
    \usepackage{epsfig}
\usepackage[final]{pdfpages}
\begin{document}

\title[Magnetic forces and
localized resonances in electron transfer through quantum rings]{Magnetic forces  and
localized resonances in electron transfer through quantum rings}

\author{M.R. Poniedzia{\l}ek and B. Szafran}

\address{Faculty of  Physics and Applied
    Computer Science, AGH University of Science and Technology, al.
    Mickiewicza 30, 30-059 Krak\'ow, Poland}
\begin{abstract}
We study the current flow through semiconductor quantum rings.
In high magnetic field the current
is usually injected to the arm of the ring preferred
by classical magnetic forces.
However, for narrow magnetic field intervals that appear periodically on the magnetic field scale
the current is injected to the other arm of the ring.
We indicate that the appearance of the anomalous -- non-classical -- current circulation results from Fano interference involving localized resonant states.
The identification of the Fano interference is based on the comparison of the solution of the scattering problem
with the results of the stabilization method. The latter employs the bound-state type calculations and allows
to extract both the energy of metastable states localized within the ring and the width of resonances by analysis
of the energy spectrum of a finite size system in function of its length.
The Fano resonances involving states of anomalous current circulation
become extremely narrow on both magnetic field and energy scales. This is consistent with the orientation
of the Lorentz force that tends to keep the electron
within the ring  and thus increases the lifetime of the electron
localization within the ring.
Absence of periodic Fano resonances in electron transfer probability
through a quantum ring containing an elastic scatterer is also explained.
\end{abstract}

\maketitle

\section{Introduction}
The Aharonov-Bohm  effect \cite{AB} consists in interference of parts of the electron wave
function passing through two paths that contain a magnetic flux inside. In the originally considered situation \cite{AB}
the magnetic field is zero within the region accessible to electrons. The experiments are usually performed in a spatially homogenous
magnetic field so the Aharonov-Bohm phase shifts are accompanied by magnetic deflection of the electron trajectories.
In particular, measurements performed for the Young double slit experiment in vacuum \cite{olariu} indicate that the envelope of the interference pattern is shifted as due to the classical Lorentz force. In solids \cite{lba} the magnetic deflection of the electron trajectories can influence the results of the interference only provided that the Larmor (cyclotron) radius  is comparable to the width of the channels.
For metal mesoscopic systems \cite{web} this condition is hardly fulfilled. In semiconductors the magnetic deflection of electron trajectories may be significant
even for the width of the channels of the order of several tens of nanometers \cite{jak}. Recently, the magnetic deflection was used in the proposal  \cite{strambinnicc}
of a solid-state setup for the interaction-free-measurement.

The effects of magnetic forces were studied for the electron injection for  semiconductor quantum point contacts \cite{qpc}
and quantum rings \cite{epl}. In two-terminal quantum rings \cite{epl} the magnetic forces lead to a preferential injection of the electron
to one of the arms of the ring (i.e. to the left arm for $B>0$). In consequence, the parts of the electron wave function that meet at the exit to the output channel are unequal which reduces the  amplitude of Aharonov-Bohm oscillations at high magnetic field ($B)$.  In three-terminal quantum rings \cite{epl} the magnetic forces besides the reduction of the Aharonov-Bohm oscillation amplitude lead to a distinct imbalance of the transfer probability to the two output terminals. An experiment on magnetic forces in three terminal quantum ring performed only recently \cite{strambini} confirmed precedent theoretical predictions \cite{epl}.
The studies of magnetic forces \cite{epl} were based on wave packet dynamics. In our previous work \cite{ps} we provided a study of magnetic forces in stationary
electron flow. The stationary solution generally reproduces the qualitative features of the wave packet dynamics simulations \cite{epl}. However,
at high magnetic field we found \cite{ps} an abrupt reversal of the current circulation and a resulting switch in the transfer probabilities
to the two output leads. It was found \cite{ps} that the interference conditions leading to the reversed current circulation appear for very narrow $B$ intervals. The physical origin of the anomalous current circulation was not explained. In the present paper we demonstrate that the reversed current circulation is due to Fano interference involving ring-localized states with magnetic dipole moment that is antiparallel to the external magnetic field.

The initial and finite states in the electron scattering problem correspond to delocalized states of the energy continuum.
The resonant scattering is qualitatively understood \cite{fano} as due to presence of a metastable state localized at the scattering object that after a finite lifetime $\tau$ decays elastically into the delocalized state. The non-stationary character of the delocalized state results in a finite width $\Gamma$ of the scattering resonance as a function of the energy. The width and lifetime are related by the uncertainty relation $\tau \Gamma\simeq\hbar$. In the present problem the initial and finite
states of the scattering process are the channel states of the lowest subband,
the energy continuum starts above the transport threshold for the lowest subband,
and the metastable states are localized within the ring.

 The Fano interference \cite{fano}
between the channel state carrying the current and
the energy-degenerate localized state of a quantum ring or dot typically
produces pronounced signatures in the electron transfer probability.
The Fano resonances are extensively studied for quantum rings and
open quantum dots side coupled to the channel \cite{clerk,
toriokondo,kang,fang, totio,lee} as well as for potential cavities
embedded within the channel \cite{gores,schmelcher}. In order to determine the
localized resonances we employ the stabilization method
\cite{stabilization0} in the version proposed by Mandelshtam and coworkers \cite{stabilization}.
The method \cite{stabilization} determines the position and width of the resonances of the scattering
probability by a bound-state calculations with square integrable eigenfunctions.
In this approach one determines the energy spectrum of a finite system in function of its size.
The spectrum contains energy levels that depend on the system size as well as size-independent energy levels. 
The size-dependent energy levels mimic
the channel eigenstates and the size-independent energy levels correspond to wave functions that are localized
at the scattering object. The latter are identified with the metastable
states giving rise to the Fano interference \cite{stabilization}.

In this paper
we present -- a first to our knowledge -- study of the interplay of
Fano resonances and magnetic forces in two- and three-terminal
quantum rings at high magnetic field.
For positive magnetic field the Lorentz force tends to inject the electron
from the incoming lead to the left arm of the ring thus inducing a clockwise current
circulation in the scattering eigenstate.
At high magnetic field
the Fano resonances involving localized states with clockwise orientation of persistent
currents become too
wide to be resolved in the transfer probability, while the
resonances with anticlockwise currents become extremely sharp.
We indicate that the changes in width of the resonances are consistent with
the lifetime of ring localized states as obtained in a time-dependent simulation.

\begin{figure}[ht!]
\begin{center}
\includegraphics[width=80mm]{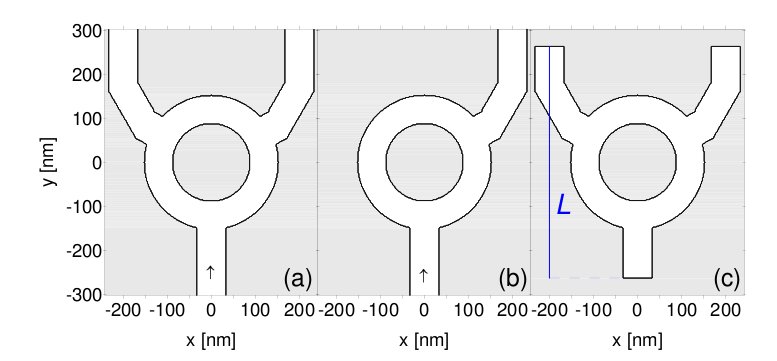}
\end{center}
\caption{Open three terminal (a) and two terminal quantum rings (b) as
well as their closed counterpart with terminals of finite width (c).
The channels are 64 nm wide, the inner and outer radii of the ring are 88 and 154 nm, respectively.
The confinement potential is assumed zero within the white area and 200 meV within the gray area.
In (a) and (b) the lead with axis $x=0$ is treated as the input channel. $L$ in (c) denotes the length of the entire closed system.
 }\label{steb}
\end{figure}

\begin{figure}[ht!]
\begin{center}
\includegraphics[width=60mm]{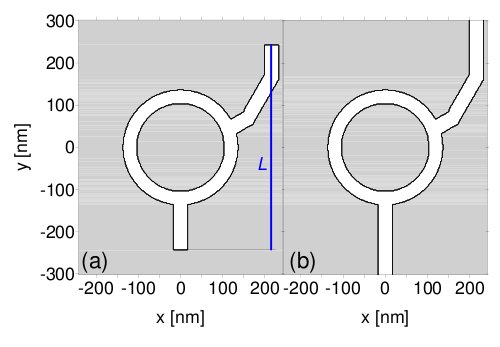}
\end{center}
\caption{ A closed quantum ring with two terminals of finite width (a)
and an open two-terminal quantum ring (b) connected to infinite channels. The channels
are 32 nm wide, the inner and outer radii of the ring are 104 and
136 nm, respectively.
 }\label{spcjaly}
\end{figure}

\section{Theory}

We study quasi two-dimensional structures in which the channels made of GaAs are embedded in Al$_x$Ga$_{1-x}$As matrix \cite{pedersen}.
We consider three- and two- terminal open quantum rings of Fig. \ref{steb}(a,b) and Fig. \ref{spcjaly}(b) as well as closed
 quantum rings with terminals of a finite length -- Fig. \ref{steb}(c) and Fig. \ref{spcjaly}(a).
 We consider systems in which the channels have 64 nm width [Fig. \ref{steb}] that were addressed in the preceding work \cite{ps}
 as well as systems with channels that are only 32 nm wide [Fig. \ref{spcjaly}] in which the effect of magnetic deflection is weaker.
 We solve the Schroedinger equation
 \begin{equation}
 H\Psi=E\Psi,
 \end{equation}
where
\begin{equation}
    H=\left({\bf p}+e{\bf A}({\bf
    r})\right)^2/{2m^*}+V(x,y),
    \label{xx}
\end{equation}
$m^*=0.067m_0$ is the GaAs effective mass and the confinement potential $V$ is taken equal to zero inside the channels and $V_0=200$ meV outside. The adopted value of $V_0$ corresponds to the Al concentration in the matrix surrounding the channels of about 27\%.
 We assume that the magnetic field is oriented perpendicular to the plane of confinement and  take the vector potential in the Landau gauge ${\bf A}=(0,Bx,0)$.

We are interested in the linear transport regime,
which corresponds to a small bias $V$ applied between the terminals. The bias raises the Fermi energy of one
 of the terminals with respect to the other and allows for a flow of an uncompensated current through the system.
The net linear current $J$ is proportional to the bias $V$, i.e.
$J(B,V)=G(B)V$, where $G(B)$ is the conductance of the ring.
In the Landauer-Buttiker approach the lowest-subband linear conductance is determined by the Fermi electron transfer probability
from the input to the output lead $G(B)=\frac{2e^2}{h}T(B)$ \cite{datta}.
The linear conductance does not depend on the bias. For that reason below we evaluate the transfer probability for $V=0$, which
is a standard approach in the linear transport calculations. In our calculation the input lead is specified by the boundary condition
and not by the bias.
The current that we calculate and discuss below corresponds to the Fermi level electron coming from the input lead.

In the lowest-subband transport regime in each of the leads far away from the ring the energy levels
are two-fold degenerate with one of the electron states going up and the other going down the channel.
In order to evaluate the transfer probability $T$ we look for solution of eigenequation (1) for the
electron that comes to the ring from the terminal that is attached to the ring from below  [Fig. 1(a)].
In the output leads far from the ring
one finds only
the outgoing wave function
\begin{equation}
\Psi(x,y)=a\exp(iqy)\psi^q(x),
\end{equation}
where $q$ is the wave vector and $\psi_q(x)$ is the transverse wave function \cite{ps}.
In the input lead one finds both the incoming and outgoing wave functions
\begin{equation}
\Psi(x,y)=b\exp(iky)\psi^k(x)+c\exp(-iky)\psi^{-k}(x).
\end{equation}
The axis of the input lead is $x=0$, while the axes of the left and right output leads [Fig. \ref{steb}] correspond to $x_l=-x_r=200$ nm.
For fixed energy the wave vectors in the input and output leads are related as $q_l=k-\frac{eB}{\hbar}x_l$ and  $q_r=k-\frac{eB}{\hbar}x_r$.
The method of evaluation of the transfer probability was described in detail in Ref. \cite{ps}. We use the kinetic energy discretization of Ref. \cite{governale}
that ensures independence of the finite-difference results of the chosen gauge.  The scattering amplitudes $a,b$ and $c$ are determined
in a self-consistent manner (see Ref. \cite{ps}).

In order to determine the energies of localized states  we solve the algebraic problem obtained in the discretized
version of the eigenequation (1) only with modified boundary conditions. We require the wave functions to vanish at the edges of computational box.
We also assume that the terminals have a finite length -- see Fig. \ref{steb}(c) and Fig. \ref{spcjaly}(a), where $L$ is the length of the system.
We determine the states localized within the ring using the stabilization method \cite{stabilization}.
When $L$ is large enough the wave function of a localized state vanishes long before the end of the channel and the energy of the ring-localized state
no longer depends on $L$. Besides the states localized in the ring, the eigenequation is solved by wave functions in which the electron is localized
inside the channels. The energies of these states decrease with $L$. The energy levels corresponding to states localized within the ring and the states of the
channel enter into avoided crossing, so one cannot indicate any fixed value of $L$ which would guarantee separation of the ring-localized states from the channel states. Instead \cite{stabilization}, the energy spectrum is calculated in function of $L$, and the energies of localized states are extracted
by a detection counter defined as
\begin{equation}
     N(E)=\int dL \sum_l \delta(|E-E_l(L)|;dE),
     \end{equation}
where the summation over $l$ runs over eigenenergies of the closed system,
\begin{equation}
\delta(|E-E_l(L)|;dE)=\left\{\begin{array}{lll}0 & \rm{for} & |E-E_l(L)| \geq dE \\ 1 & \rm{for} & |E-E_l(L)| < dE\end{array}\right.,
\end{equation}
and $dE$ is a small energy window. In the calculation we keep a constant mesh spacing of 2 nm and  $L$ takes
 discrete values that are integer multiples of 4 nm. We consider $L$ between 480 and 600 nm. Under these assumptions Eq. (5) amounts in counting the energy levels that appear close to $E$ as $L$ is changed.

\begin{figure}[ht!]
\begin{center}
\begin{tabular}{ccc}
\includegraphics[width=30mm]{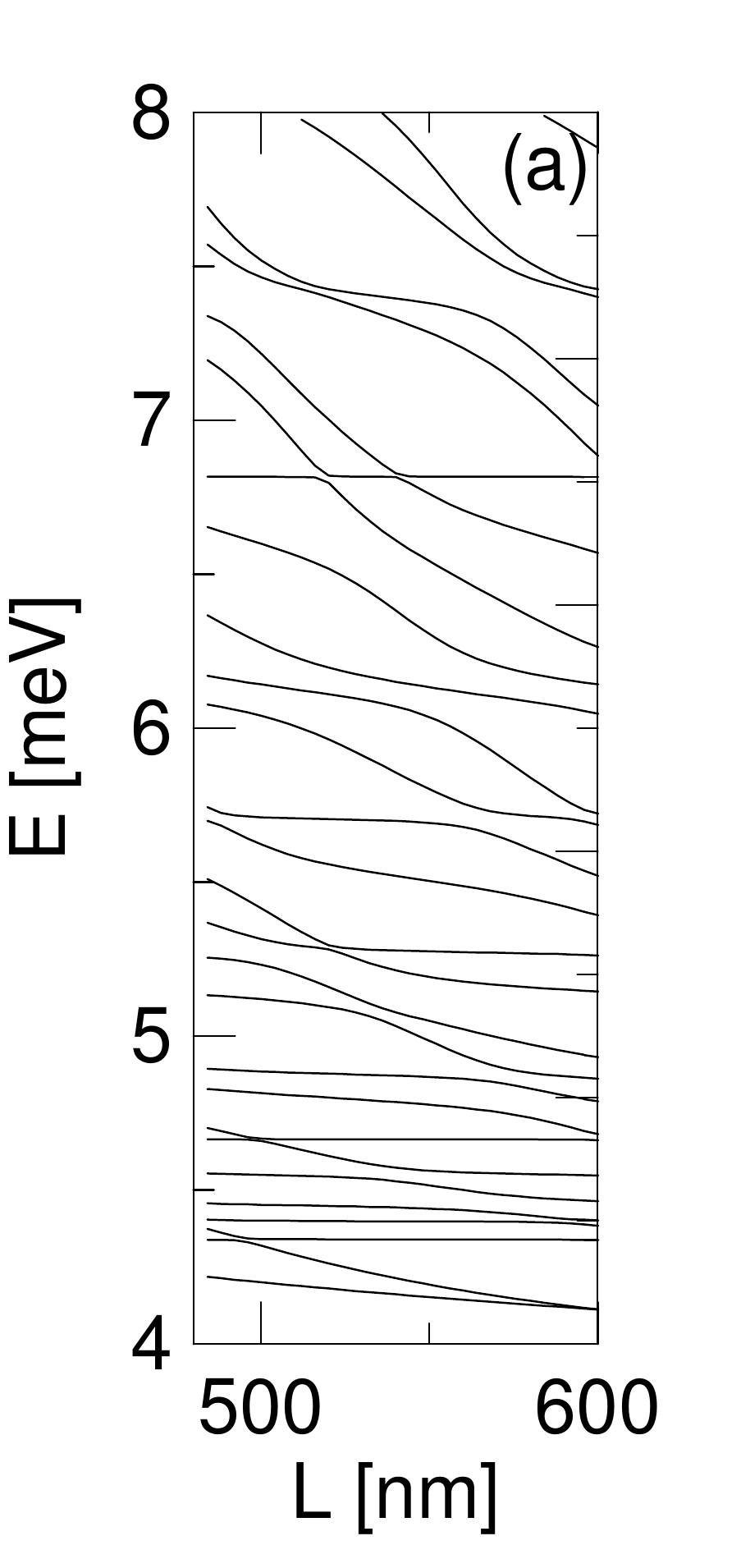} &
\includegraphics[width=30mm]{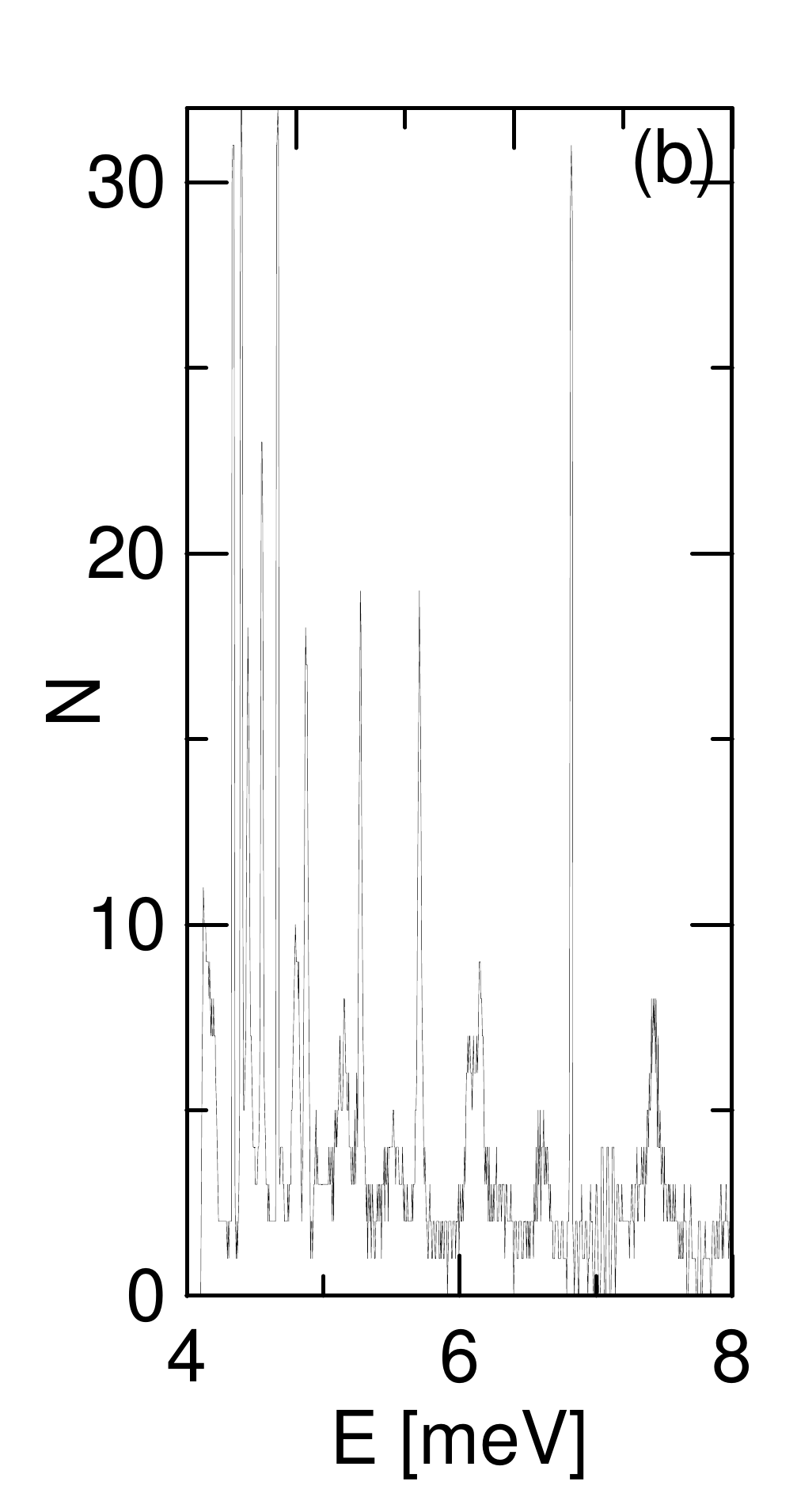} &
\includegraphics[width=50mm]{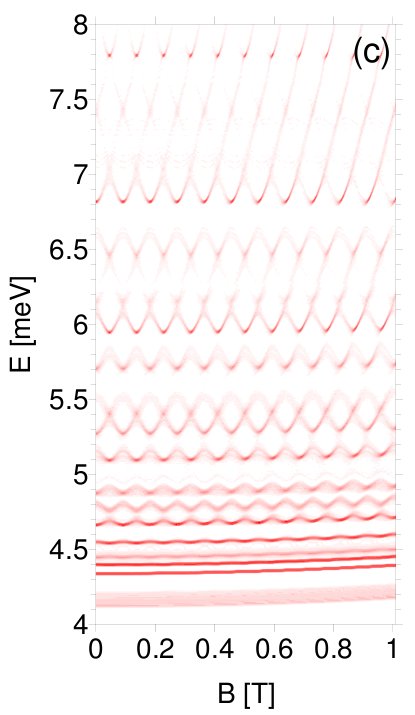}
\end{tabular}
\end{center}
\caption{(a) Energy spectrum of a closed ring with two terminals
attached [Fig. \ref{spcjaly}(a)] as a function of the length of the
system for $B=0$. (b) The localized states  detection
counter $N$ [Eq. (5)] for $B=0$ obtained for the energy window $dE=10\mu$eV. (c) The resonance detection counter in function of $E$ and $B$. The darker
the shade of red the larger value of $N$.
}\label{closeds}
\end{figure}

\begin{figure}[ht!]
\begin{center}
\includegraphics[width=100mm]{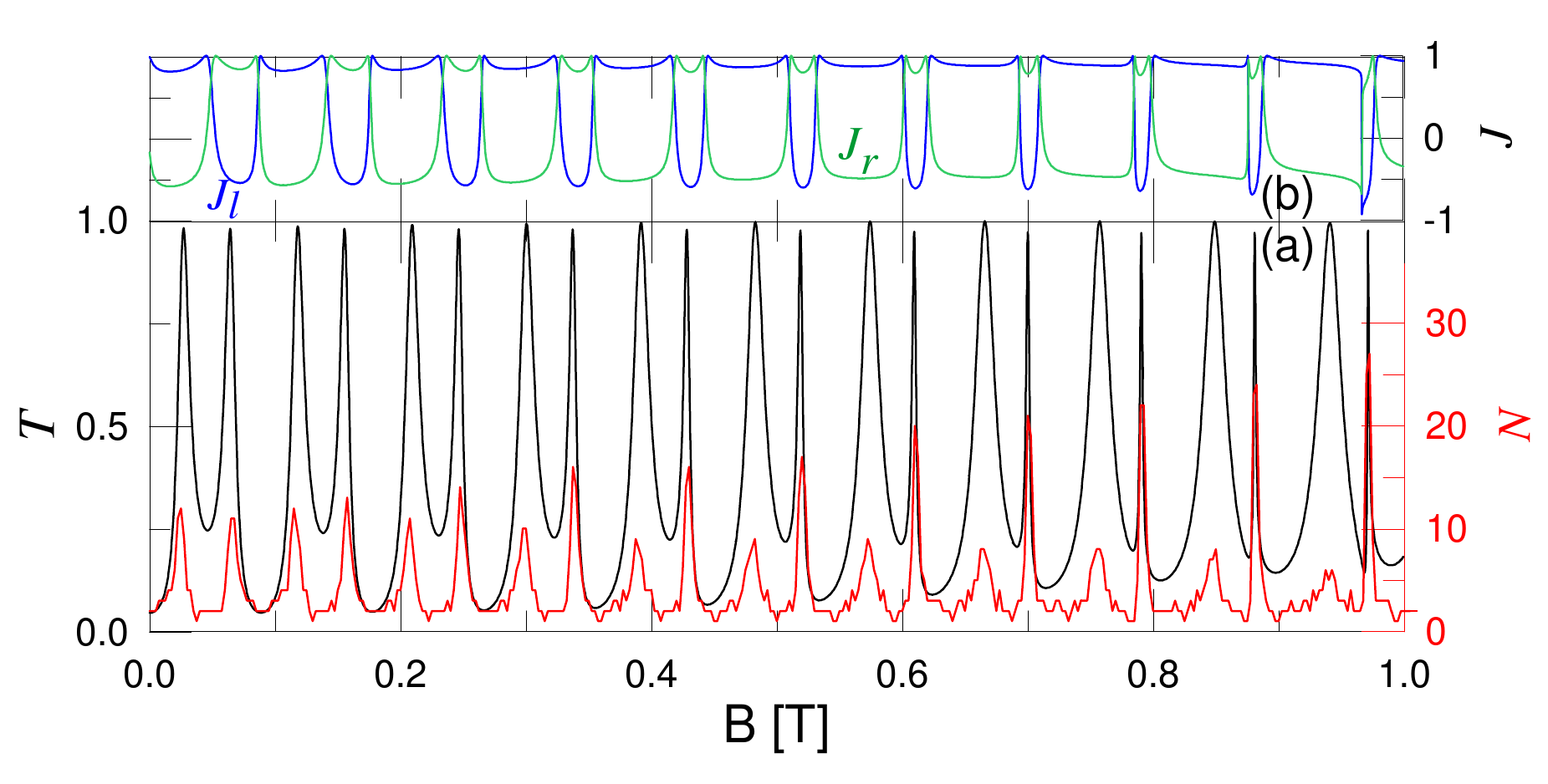} 
\end{center}
\caption{(a) The black curve shows the electron transfer probability (left axis) through the
two terminal ring of Fig. \ref{spcjaly} for the incident electron energy of 6 meV  and the red curve presents the localized resonance detection counter (right axis) -- a cross section of Fig. \ref{closeds}(c). (b) The normalized density current fluxes through the left $J_l$ and right $J_r$ arms of the ring calculated for $y=0$ (see Fig. 2).
 }\label{wryski}
\end{figure}

\begin{figure}[ht!]
\begin{center}
\includegraphics[width=40mm]{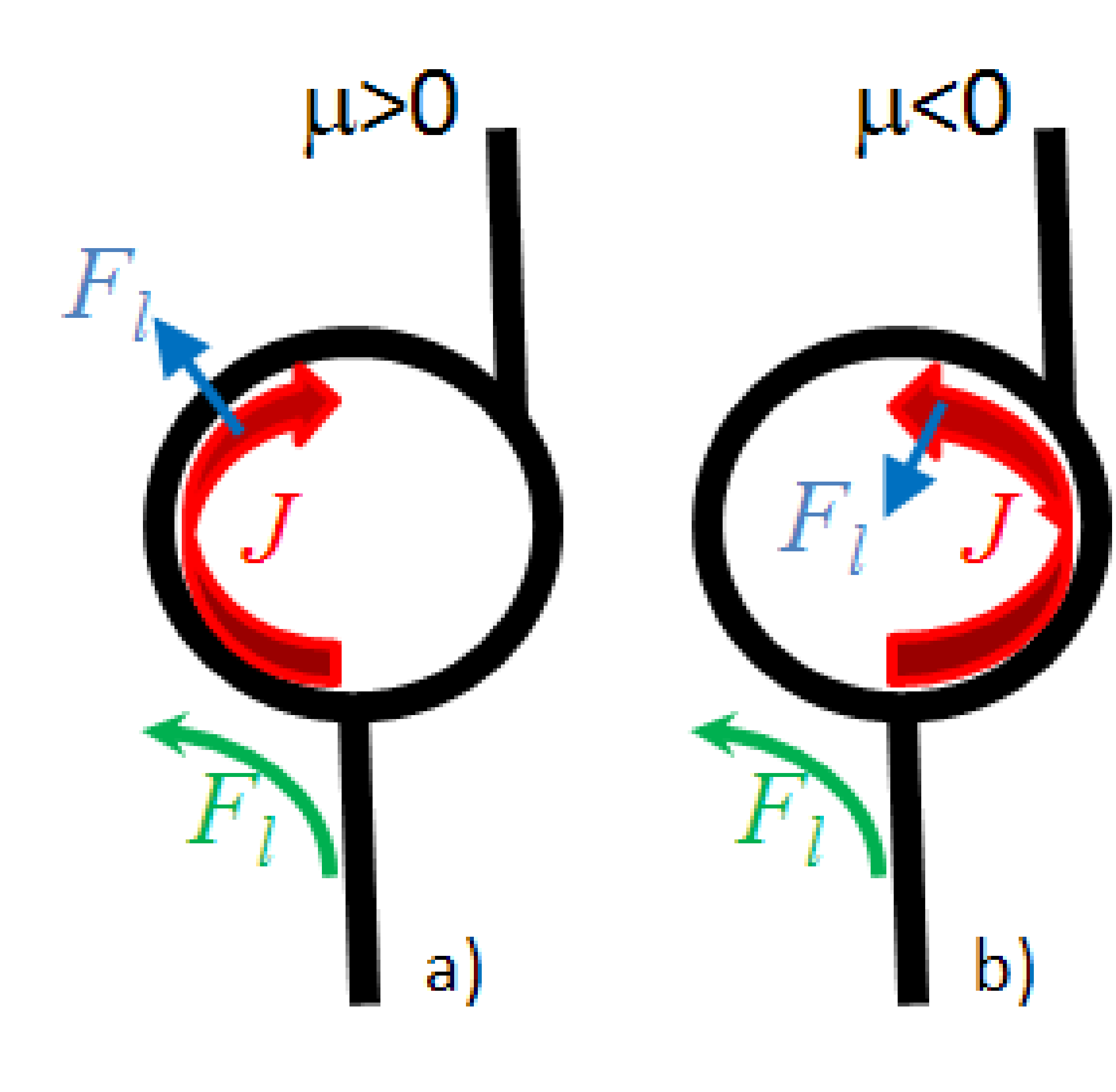} 
\end{center}
\caption{Schematic drawing of the probability current circulation $J$ (red arrows) in the resonant states corresponding to the positive (a) and negative (b) magnetic dipole moment $\mu$ at $B>0$.
The green arrows indicate the Lorentz force that tends to inject the electron to the left arm of the ring.
The blue arrows indicate the direction of the Lorentz force acting on the electron that circulates within the ring.
For the circulation producing  $\mu>0$  ($\mu<0$) the Lorentz force tends to eject the electron to the leads (keep the electron within the ring).
 }\label{sche}
\end{figure}

\begin{figure}[ht!]
\begin{center}
\includegraphics[width=100mm]{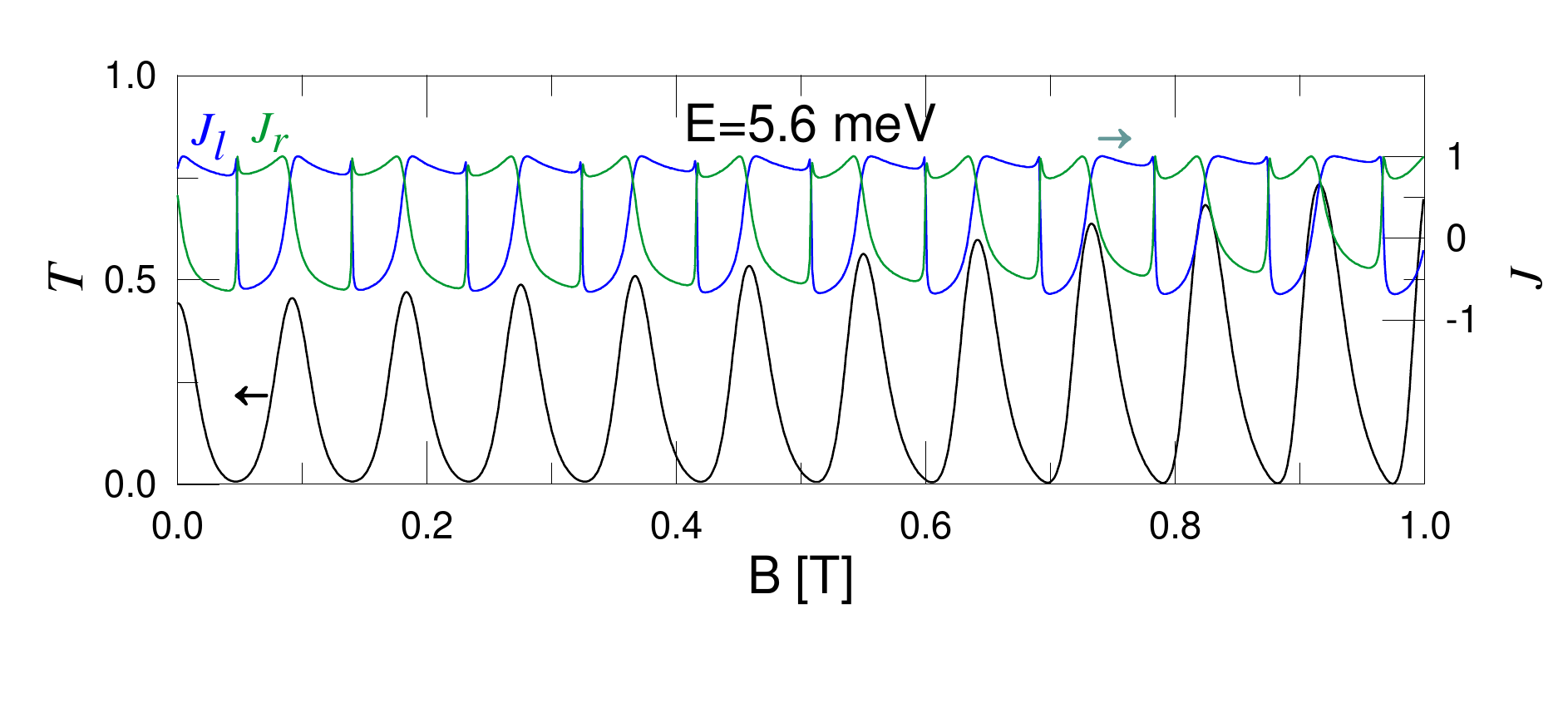} 
\end{center}
\caption{Same as Fig. \ref{wryski} but for the incident electron of energy 5.6  meV.}\label{wryskino}
\end{figure}

\section{Results and discussion}
\subsection{Thin channels}
For the clarity of the presentation it is useful to start from the ring with two terminals and thin channels of Fig. \ref{spcjaly}.
We first extract the localized states by the stabilization method. We consider a closed system of a finite length $L$
that is depicted in Fig. \ref{spcjaly}(a).
The eigenenergies found as function of $L$ are plotted in Fig. \ref{closeds}(a) for $B=0$.
All the energies presented in Fig. \ref{closeds}(a) correspond to the energy continuum.
For the considered channel width the continuum threshold
is located at 3.89 meV at $B=0$  (3.95 meV at $B=1$ T).
The energy levels
that are independent of $L$ correspond to states with wave functions localized within the ring.
The other energy levels correspond to delocalized states with wave functions that vanish only at the end of the channels.
In Fig. \ref{closeds}(b) we displayed the localized states detection counter $N$ [Eq. (5)] calculated from the spectrum of Fig. \ref{closeds}(a) according to formula Eq. (5) with the energy window $dE=10\mu$eV. The counter $N$ exhibits sharp peaks for energies for which a presence of a localized state is evident -- see the flat energy level near 6.8 meV in Fig. \ref{closeds}(a). $N$ forms wider maxima when a presence of localized state is less evident, in particular, where it is only suggested by a wider avoided crossings in Fig. \ref{closeds}(a) -- see for instance the maximum of $N$ found near 6.1 meV.

In Fig. \ref{closeds}(c) we plotted how the maxima of $N$ move along the energy scale as the external magnetic field is applied. The plot exhibits a clear Aharonov-Bohm oscillation
of the resonances
with a period of about 0.091 T,
which corresponds to the magnetic field that produces a quantum of the flux $\Phi_0=h/e=B\pi R^2$ piercing a one-dimensional ring of an effective radius $R=120$ nm, which well agrees with the applied geometry (the arithmetic average of the inner and outer radii of the ring is 121 nm). The resonant energy levels increase on average with $B$ due to the diamagnetic shift resulting from the finite width of the channels. For a closed perfectly circular quantum ring the angular momentum is a good quantum number and the energy levels cross. The avoided crossing that are present in Fig. 3(c) are due to angular momentum mixing by perturbation of the circular symmetry that is introduced by attachment of the input and output terminals [see Fig. \ref{spcjaly}].

In Fig. \ref{wryski}(a) we plotted with the red curve the values of $N$ found for  $E=6$ meV in function of $B$. The result is obtained as a cross section of Fig. \ref{closeds}(c) for a fixed energy value.  Within a magnetic field range of 0.091 T corresponding to a single flux quantum, we observe
a double peak structure in the $N(B)$ dependence. As $B$ grows one of the peaks of the pair becomes sharper, and the other is transformed into  lower and wider maximum.
The sharp (wide) peaks of Fig. \ref{wryski}(a) correspond to localized energy levels that grow (decrease) in energy as $B$ grows
[cf. Fig. \ref{closeds}(c)].

Let us now consider the open system [Fig. \ref{spcjaly}(b)] and an electron of energy 6 meV that comes to the ring from below. The calculated transfer probability
is displayed in Fig. \ref{wryski}(a) with the black line. We see that the transfer probability possesses peaks as functions of $B$ which coincide
with the maxima of the resonance detection counter $N$.
The result of Fig. \ref{wryski}(a) indicates that the transfer probability is governed by the interference of the channel state with a localized state, known as the Fano \cite{fano} interference. The transfer probability extrema are distinctly asymmetric which is a typical signature of this phenomenon \cite{fano}.

Note, that in Fig. \ref{wryski}(a) a good agreement is obtained not only in
the position but also in the width of $T$ and $N$ maxima.
We find that the maxima of $T$ and $N$ become wider or thinner at higher $B$ depending on the direction of the persistent current circulation in the localized
states.
For a closed quantum ring the magnetic dipole moment generated by the persistent current flowing in the stationary state $\psi$ is defined as $\mu=-\frac{d E}{dB}=-\frac{d\langle \psi |H|\psi \rangle }{dB}$. For a strictly one-dimensional ring one obtains by an elementary algebra \cite{szafranes}
a classical formula $\mu=-\frac{1}{2}e {\bf r} \times {\bf j}$, where ${\bf j}$ is the probability density current.
The states whose energy increases (decreases) with growing $B$ produce the magnetic dipole moment which is antiparallel (parallel) to the external magnetic field
with anticlockwise (clockwise) probability density current circulation around the ring (see Fig. \ref{sche}).

In Fig. \ref{wryski}(b) we plotted the normalized probability density current fluxes calculated across the left and right arms of the ring
across the line $y=0$.
Note, that between adjacent maxima of $T$ and $N$ the circulation of the current changes orientation.
The maxima of $T$ that become wider at high $B$ correspond to clockwise current circulation  ($J_r<0, J_l>0$,
i.e. the current going down in the right arm and up in the left arm).
The clockwise current circulation is consistent with the orientation of the current flow in the localized resonance that
enters into degeneracy with the incident electron energy in the resonant range of the magnetic field. Moreover,
the clockwise current agrees with the  orientation of magnetic forces which tend to inject the electron from the input channel
to the left arm of the ring (see the sketch of Fig. \ref{sche}). The intervals of clockwise current circulation increase in width at higher $B$
and so do the corresponding Fano resonances of the transfer probability.
The Fano resonances that become extremely thin at high $B$ involve localized states with anticlockwise current circulation that is opposite
to the direction of the Lorentz force at the electron injection to the ring.

Fig. \ref{wryski}(b) contained a number of Fano peaks of the transfer probability obtained for the electron energy of 6 meV.
Fig. \ref{wryskino} shows the transfer probability calculated for
the incident electron energy of 5.6 meV, in the center of avoided
crossings between the localized states [see Fig. \ref{closeds}(c)].
For this energy there are no conditions for Fano interference, i.e.
at the energy of the incident electron no localized energy level is
found for any magnetic field. Fig. \ref{wryskino} indicates that the
transfer probability undergoes Aharonov-Bohm oscillation with no
abrupt features present at high magnetic field. For the transfer
through the localized states [Fig. \ref{wryski}] the peaks of $T$
corresponded to a distinct clockwise or anticlockwise current
circulation. When the Fano interference is absent [Fig. \ref{wryskino}] -- at the peaks of
$T$  -- the current flows with the same intensity through both arms
of the ring in spite of the asymmetry in the
attachment of the leads to the ring [Fig. \ref{spcjaly}] accompanied by
asymmetry of the magnetic forces. The effect of the magnetic
injection is still visible in Fig. \ref{wryskino}:  the magnetic
field windows that correspond to clockwise current circulation
become wider at higher $B$ at the expense of these $B$ ranges in which
the current flows anticlockwise.

\begin{figure}[ht!]
\begin{center}
\includegraphics[width=120mm]{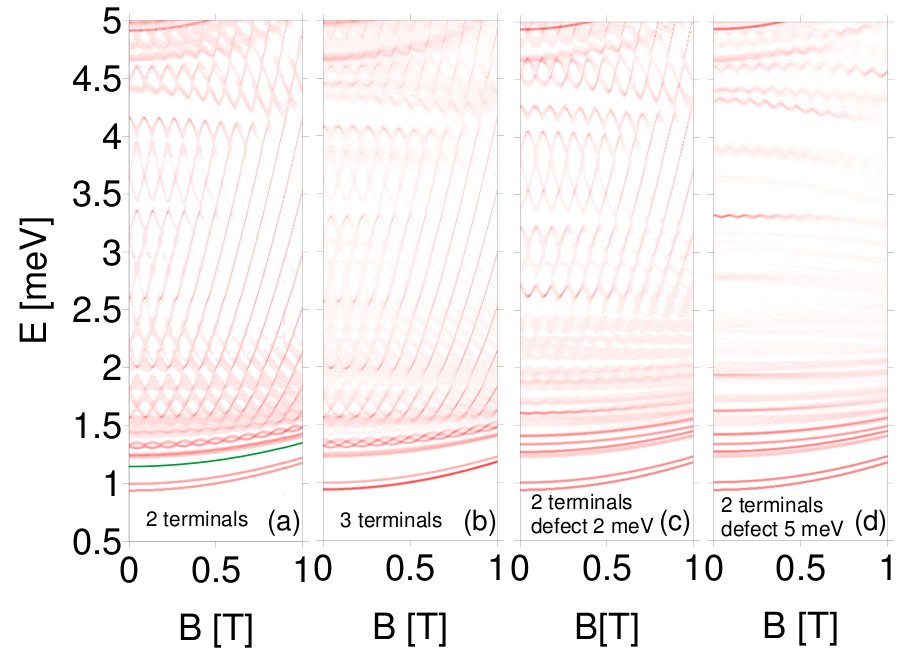} 
\end{center}
\caption{Resonance detection counter $N$ [see Eq. (5)] for two-terminal (a) and  three-terminal (b) perfect ring. The darker
the shade of red the larger the value of $N$.
The results for a two-terminal ring with a repulsive Gaussian defect (c,d) present with the ring (see text).
The height of the defect is taken equal to 2 meV in (c) and 5 meV in (d). In (a) by the green curve we plotted the energy of
the continuum threshold.
 }\label{stab}
\end{figure}

\begin{figure}[ht!]
\begin{center}
\includegraphics[width=120mm]{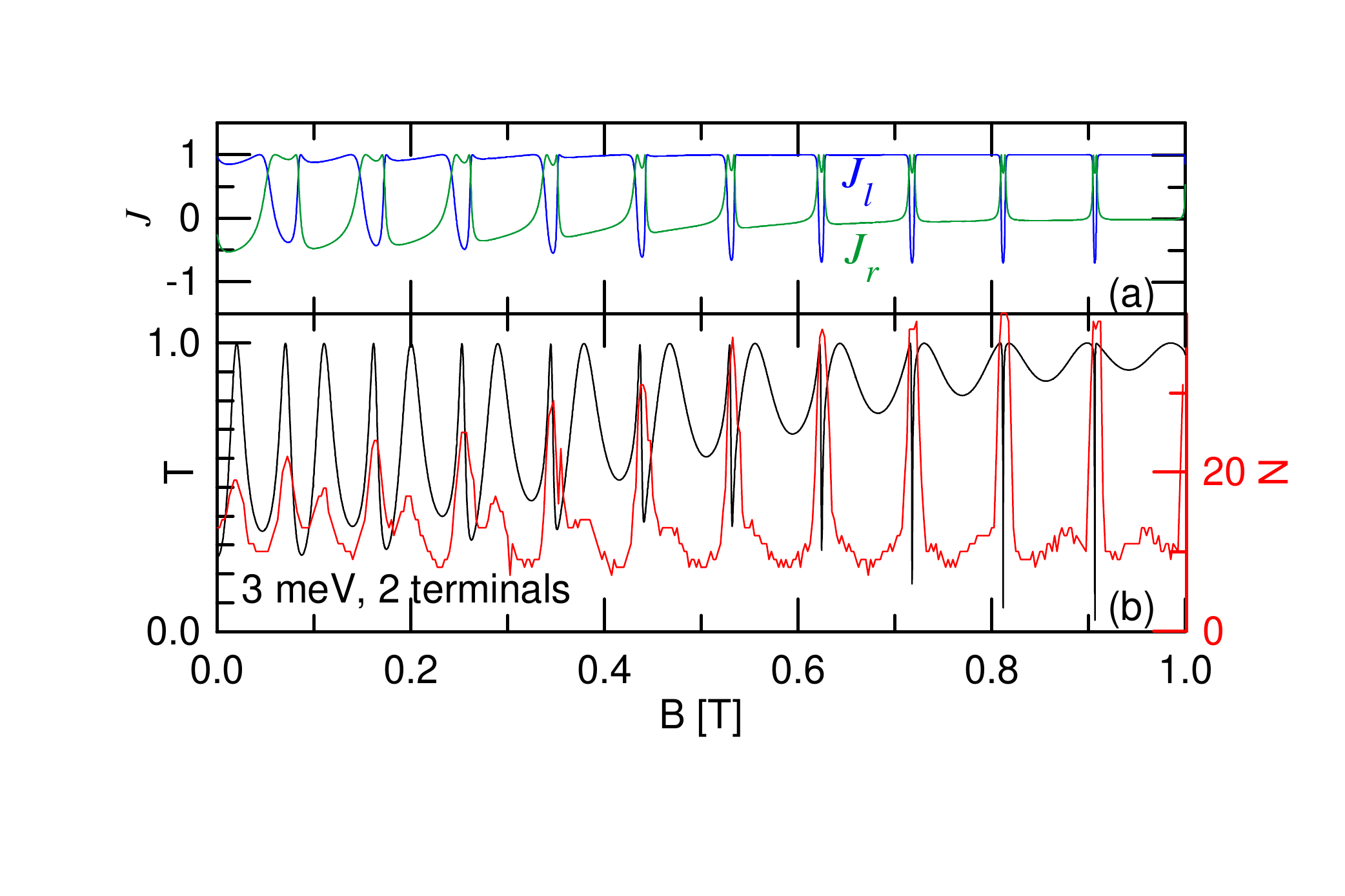} 
\end{center}
\caption{Electron transfer probability [black curve  in (b) referred to the left axis] and resonance detection counter [red curve in (b) referred to the right axis] as functions
of the magnetic field for a two terminal ring [see Fig. \ref{steb}(b)]. The green and blue curves in (a) show the normalized current fluxes through the
left and right arm of the ring calculated for $y=0$. }\label{trzye}
\end{figure}

\begin{figure}[ht!]
\begin{center}
\includegraphics[width=100mm]{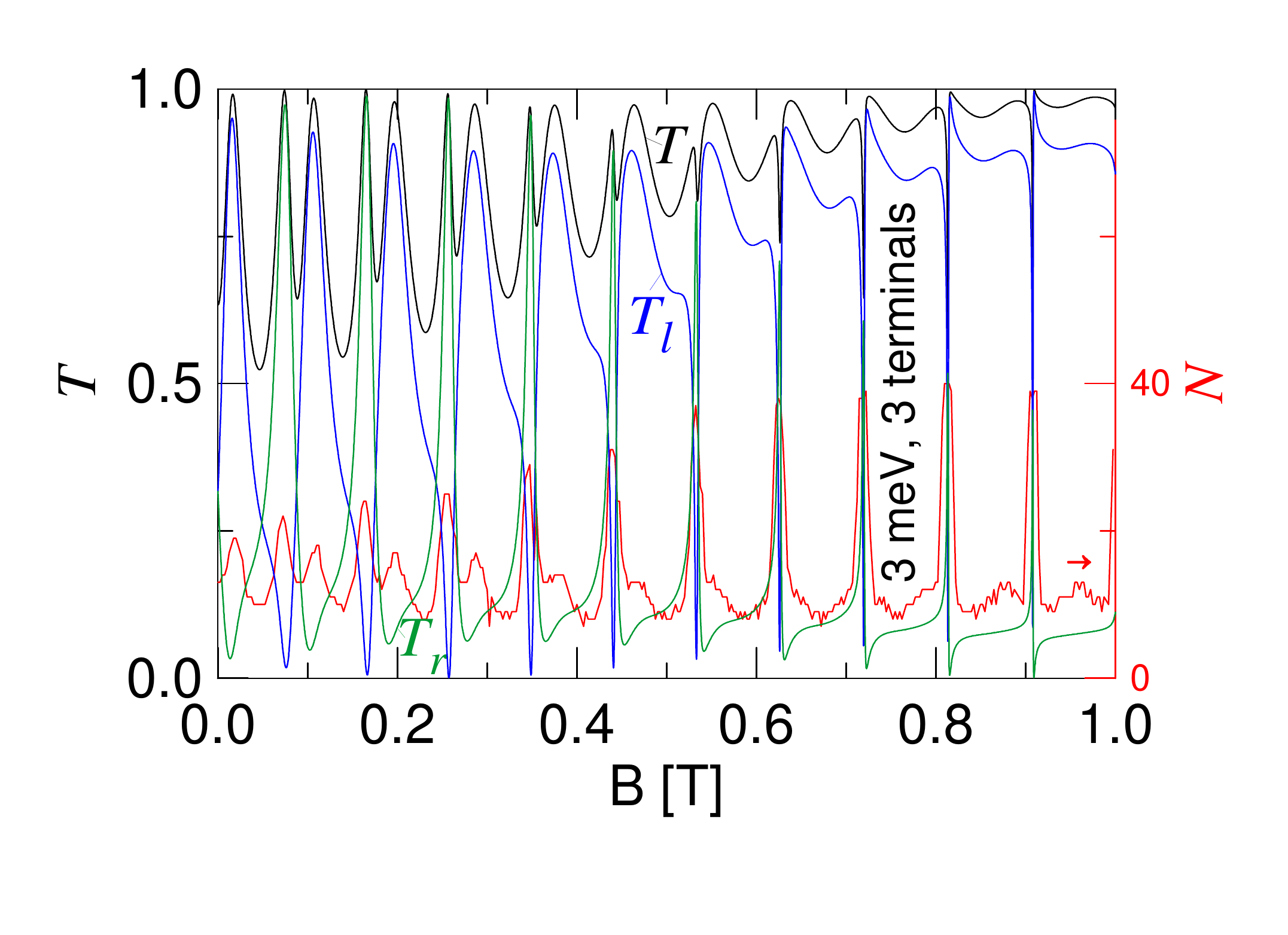}

\end{center}
\caption{Electron transfer probability [$T=T_l+T_r$, the black curve, the left axis] and resonance detection counter [the red curve, the right axis] as functions
of the magnetic field for a three terminal ring [see Fig. \ref{steb}(a)].
The green and blue curves in (a) show the transfer probability to the left and right output leads. The incident electron energy of 3 meV is applied as in Fig. \ref{trzye}.}\label{trzytrzyk}
\end{figure}

\begin{figure}[ht!]
\begin{center}
\begin{tabular}{c}
\includegraphics[width=80mm]{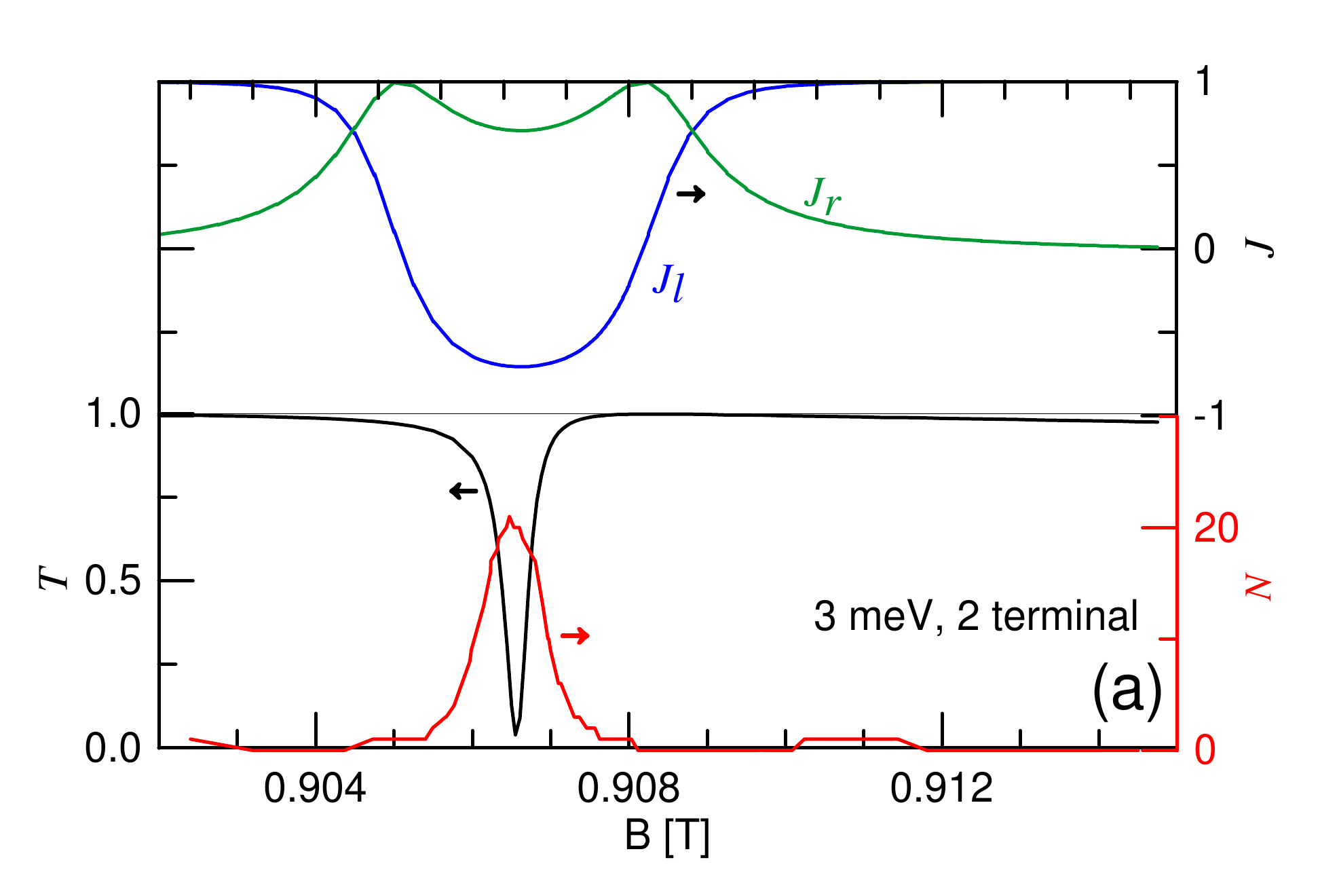} \\
\includegraphics[width=80mm]{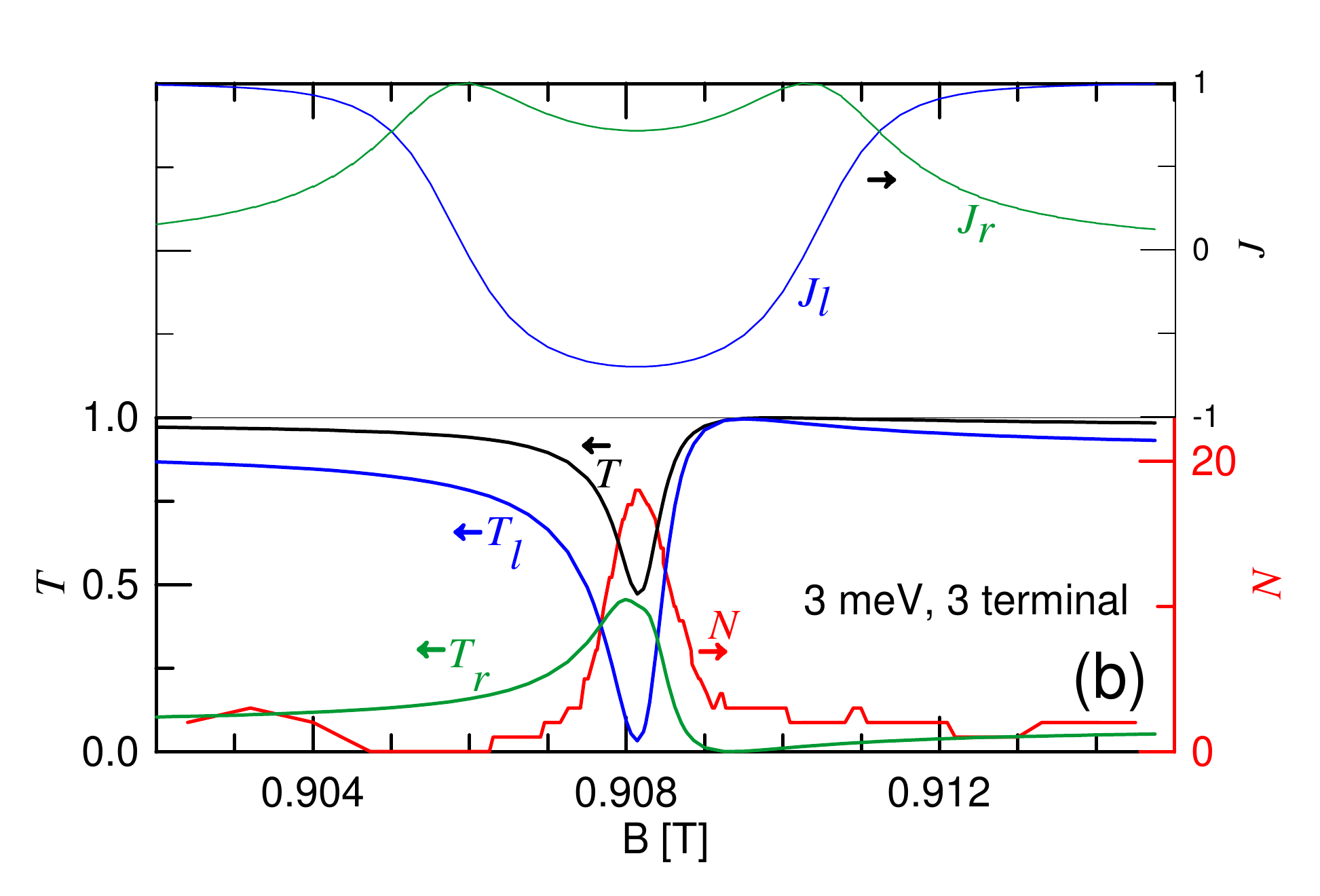}
\end{tabular}
\end{center}
\caption{(a) and (b) show enlarged fragments of Fig. \ref{trzye} and Fig. \ref{trzytrzyk}, respectively.
In this plot the values of $N(B)$ were obtained for $dE=2\mu$eV.}\label{zoomy}
\end{figure}

\subsection{Wider channels}

Now let us turn our attention to the ring with wider channels [Fig.
\ref{steb}] in which the deflection of the electron trajectories by the Lorentz force
is much stronger. The resonance detector counter as calculated for two
[Fig. \ref{steb}(b)] and three terminal [Fig. \ref{steb}(b)] device
is displayed in Figs. \ref{stab}(a) and \ref{stab}(b), respectively.
The green curve in Fig. \ref{stab}(a) displays the position of the
continuum threshold, determined as the ground-state of an infinite
channel that corresponds to zero wave vector. The two lowest energy
lines correspond to bound states of which the ground-state is
localized at the junction of the output terminal to the ring  and
the higher energy state -- at the input terminal to the ring. Above
the threshold one observes ring-localized resonant levels. As compared
to the case with thinner channels presented in Fig. \ref{closeds}(c)
the energy of the ring localized states is shifted down by about 3
meV on the energy scale and the avoided crossings between the
resonances are much thinner. The spectrum of resonant energy levels
in two terminal [Fig. \ref{stab}(a)] and three terminal [Fig.
\ref{stab}(b)] quantum rings are very similar.

In Fig. \ref{trzye}(b) we plotted the transfer probability (black line) and resonance detection counter (red line)
for the incident electron energy of 3 meV.
We can see that the double peak structure of $T$ evolves in the magnetic field much faster than for thinner channels
[Fig. \ref{wryski}]. In Fig. \ref{trzye}(a) we plotted the fluxes of the current through both arms of the ring. As previously [Fig. \ref{wryski}]
the $T$ peaks with clockwise current circulation
widen in higher $B$. The peaks of $T$ that are associated with current circulation that is anticlockwise -- opposite to the magnetic injection -- become very narrow at higher $B$. At higher $B$ the resonance detection counter resolves only the narrow Fano peaks. We can see that within a single $N$ maximum
at higher $B$ one obtains a peak that is followed by an abrupt dip. With the exception of the abrupt dips the transfer probability becomes close to 1 and the oscillations acquire the Aharonov-Bohm periodicity with a decreasing oscillation amplitude.

The result for three terminal ring is displayed in Fig. \ref{trzytrzyk}. Each of the double peaks of $T=T_l+T_r$ correspond to transfer to the left ($T_l$) or right ($T_r$) output electrodes. $T$ for the three terminal ring is similar to the one obtained for two terminal structure [Fig. \ref{trzye}]. The dips of $T$ are associated with peaks of the transfer probability to the right output terminal $T_r$ and dips of $T_l$.

In Fig. \ref{zoomy} we plotted a zoom of the data that corresponds to the dip of $T$ near 0.91 T for both the two-terminal [Fig. \ref{zoomy}(a)] and the three-terminal rings [Fig. \ref{zoomy}(b)].
The dips of $T$ are distinctly asymmetric and coincide with the peak of $N$ as calculated for $dE=2\mu$eV. For three terminal ring [Fig.
\ref{zoomy}(b)] an asymmetric peak of $T_r$ coincides with an asymmetric dip of $T_l$.

\subsection{Lifetime of ring localized states versus the magnetic forces}
 Figs. 3(a) and 7 indicate that at high magnetic field  the resonances corresponding to localized states with $\mu<0$  ($\mu>0$) decrease (increase) in width as a function of both $B$ and $E$.
The width of the Fano resonance as a function of the energy is related to the lifetime \cite{fano}  of the corresponding localized state.
In order to study the lifetime of ring-localized states we solved the time-dependent Schroedinger equation with the method described in Ref. \cite{ps}. For the initial condition we took the eigenstates of a closed ring with no leads attached [see Figs. 11(b) and Figs. 11(e)] which in this simulation play the role of the metastable states of the open system.
We focused
on the magnetic field near 0.91 T and on the electron energy of 3 meV which was considered in the last subsection.
The electron forms eigenstates of this energy at 0.905 T and 0.915 T in the closed ring [Fig. 11(a)].
The eigenstates possess a definite angular momentum of $-6\hbar$ and $-17\hbar$, respectively. The $l=-6$ state corresponds to $\mu<0$,
the Lorentz force tends to keep the electron within the ring [Fig. \ref{sche}(b)] and the charge density is pushed into the inner edge of the ring
[Fig. 11(b)]. Opposite shift is observed for $l=-17$ which near 0.9 T corresponds to $\mu>0$, see Fig. 11(e).

In the initial moment of the simulation ($t=0$) we attach the leads to the ring. For $l=-17$ the packet is promptly ejected to the channel [Fig. 11(f,g)], while for $l=-6$ the Lorentz force keeps the electron
within the ring [Fig. 11(c,d)]. The part of the wave packet that is localized within the ring for both initial conditions are displayed
in Fig. 12 with the red ($l=-17$) and the black ($l=-6$) lines. For $l=-6$ the ring-localized part falls to 50\% near 100 ps as compared to  4.2 ps for $l=-17$.

For comparison we also performed simulations for eigenstates of the energy equal to 3 meV at lower magnetic fields $\simeq 0.1$ T,
where the Lorentz force is weaker.
We took $l=6$  and $l=-8$ closed-ring eigenstates for the initial condition, for the corresponding magnetic fields of 0.0625 T and 0.12 T.
Fig. 12 shows that the decay of the probability to find the electron
within the ring for these two initial states of opposite current circulation is similar.

\subsection{Rings with elastic scatterers}
The experimental data \cite{strambini} for the three terminal quantum ring
exhibit the conductance asymmetry to the two output leads at $B=0$, accompanied by a low visibility of conductance oscillations at low magnetic field.
In Ref. \cite{posza} we demonstrated by wave packet simulations that both the 0T conductance asymmetry and low visibility of the oscillations
can be explained by elastic scattering effects due to a repulsive defect present within the ring.
No sharp features of the transfer probability that might be due to the Fano interference were observed in the experimental results \cite{strambini}.
A possible reason for their absence is the finite temperature effect since the sharp features due to the Fano resonances are thermally unstable \cite{ps}. However, in Ref. \cite{ps} (Fig. 14 in particular)
we demonstrated that the presence of the elastic scatterer explaining the 0T conductance asymmetry and low visibility of
conductance oscillations  implies removal of any sharp features of the transfer probability at high magnetic field
{\it also at 0K}.
The present study of the localized states explains this finding.

In order to establish the role of elastic scatterers for the localized resonances we considered the ring with wide channels and
two terminals of Fig. 1(b). As an elastic scatterer we used a Gaussian potential defect \cite{ps,posza} $W(x,y)=V_0 \exp(-((x-x_c)^2+(y-y_c)^2)/R^2)$ where the center of the defect $x_c,y_c$ is put in point (-104,-60) nm and its diameter $2R=64$ nm is taken equal to the width of the channel. The results for the localized states detection counter
is displayed in Figs. 7(c) and 7(d) for the height of the defect of 2 meV and 5 meV.  In Fig. 7(c) we notice that the resonances
of the energy lower than 2.5 meV no longer exhibit Aharonov-Bohm periodicity. An electron with lower energy no longer tunnels across the defect and
the circulation of the persistent current along the rings is significantly hampered. The confinement potential for these electron energies is of a bent quantum wire type rather than a quantum ring.
In Fig. 7(d) we can see that the Aharonov-Bohm periodicity is removed of the energy spectrum below 4 meV. The resonances still exist, but their energies are  nearly flat in function of $B$, and they only exhibit a diamagnetic shift. For a quantum ring without the defect [Fig. 7(a,b)] at a fixed value of the Fermi energy the resonances appear with the Aharonov-Bohm periodicity. When the elastic scatterer is present the resonances appear at a sweep of $B$ only for certain energies, and their appearance -- if any -- can only be occasional and not periodic in $B$, which explains the removal of sharp Fano resonances of the theoretical spectra \cite{ps} and is a likely reason for their absence also in the experimental data \cite{strambini}.

\begin{figure*}[ht!]
\begin{center}
\includegraphics[width=140mm]{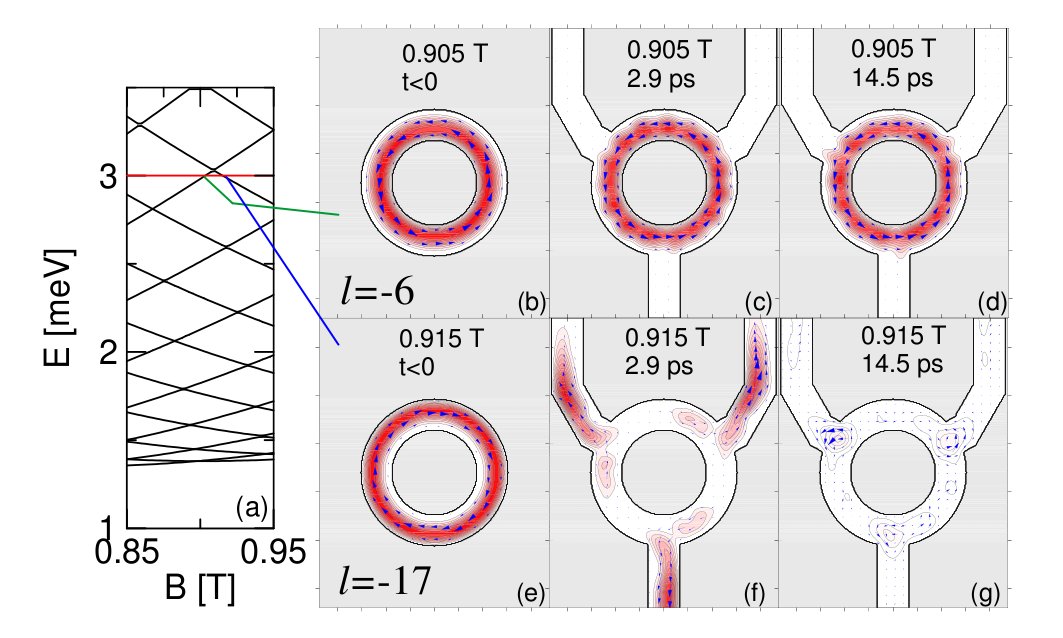}
\end{center}
\caption{(a) Energy spectrum of a closed ring. (b-g) Snapshots of the time-dependent simulations for the initial condition
taken as the eigenstates of a closed ring of the energy of 3 meV, that for $B=0.905$ T corresponds to angular momentum quantum
number $l=-6$ (b) and for $B=0.915$ T to $l=-17$ (e). (b) and (e) show the initial condition (eigenfunctions of a closed ring).
The contours show the charge density and the blue arrows - the current density distribution.
(c) and (d) show the results  $2.9$ ps after the leads are attached to the ring.
The snapshots corresponding to $t=14.5$ ps are displayed in (d) and (g).
 }\label{time1}
\end{figure*}

\begin{figure}[ht!]
\begin{center}
\includegraphics[width=70mm]{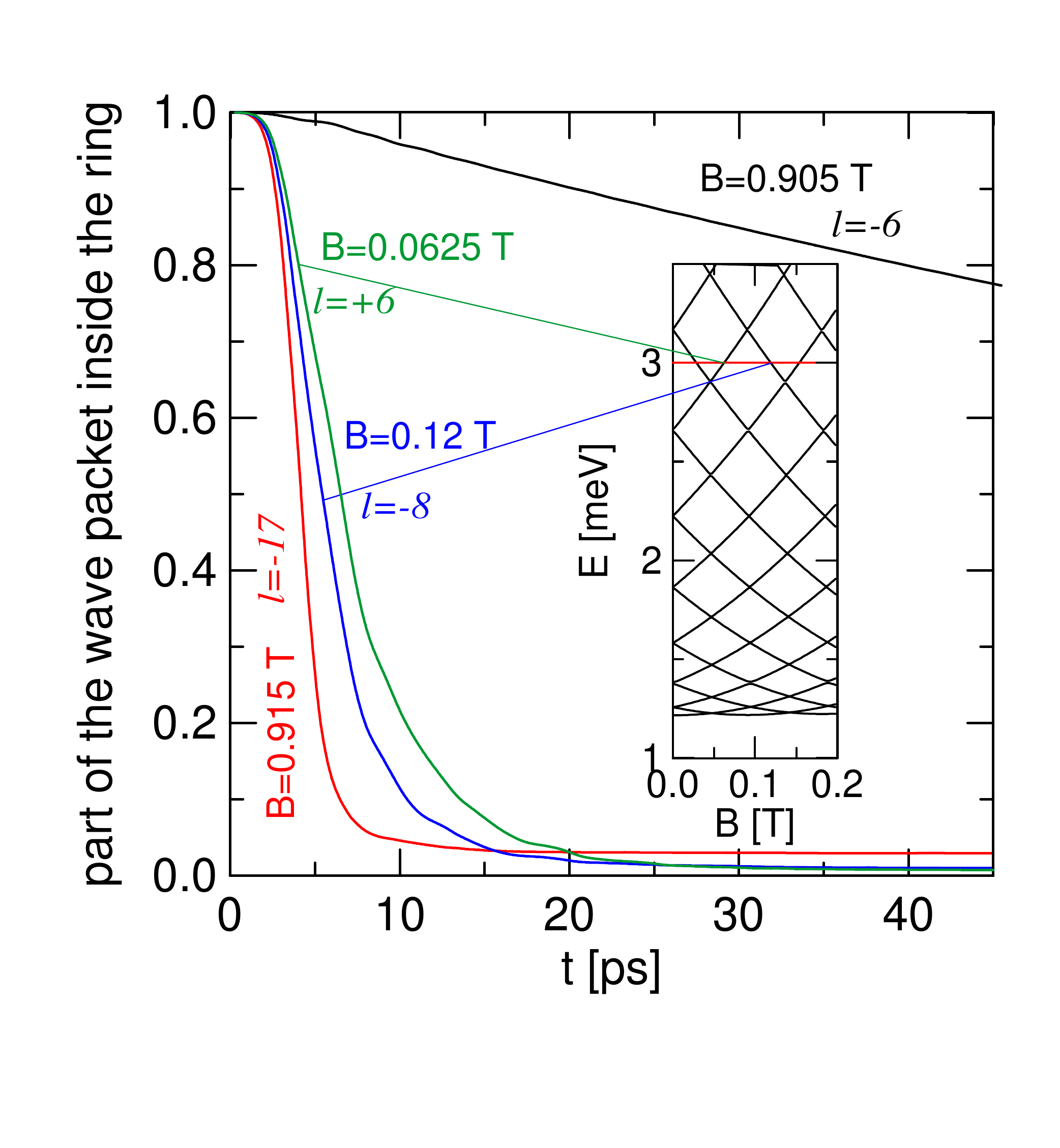}
\end{center}
\caption{The red and black curves show the part of the wave packet localized
within the ring for the time-dependent simulations presented in Fig. \ref{time1}.
The blue and green curves show the results of the simulation obtained for lower magnetic fields
for $l=-8$ and $l=6$ angular momenta and the energy of 3 meV.
Inset shows the closed ring spectrum for low magnetic fields. }
\label{time2}
\end{figure}

\section{Summary and Conclusions}

We studied the
electron transport through a two- and three-terminal quantum rings in external magnetic field.
We demonstrated that at low magnetic field the
electron transfer probability bears signatures of Fano interference
with localized states of both clockwise and anticlockwise persistent
current orientation, which produce magnetic dipole moment that is
parallel and antiparallel to the external magnetic field,
respectively. In the localized states producing the magnetic dipole moment
that is
antiparallel to the magnetic field vector the Lorentz force
tends to keep the electron within the ring thus
enhancing the lifetime of the ring localized states.
The corresponding
resonances in the transfer probability dependence at high magnetic
field become very narrow in function of both $B$ and the energy.
Opposite is the
behavior of Fano resonances involving localized states with
current circulation which produces the magnetic dipole parallel to the external magnetic field:
at high $B$ they become too wide to be
resolved in the dependence of the transfer probability on the
magnetic field.
This current circulation (clockwise for $B>0$) is consistent with the direction
of the magnetic-field-assisted electron injection to the ring (to the left arm of the ring for $B>0$).
We also demonstrated that an elastic scatterer
present within the ring destroys the Aharonov-Bohm periodicity of
the Fano resonances.

    {\bf Acknowledgements}
This work was performed within a research project N N202 103938
supported by Ministry of Science an Higher Education (MNiSW) for 2010-2013.
 Calculations were performed in
    ACK\---CY\-F\-RO\-NET\---AGH on the RackServer Zeus.


\begin{thebibliography}{00}
\bibitem{jak} The Larmor radius is $R_L=m^*V_F/eB$ where $m^*$ is the effective mass and $V_F$ is the Fermi velocity. The semiconductor effective mass and the Fermi velocity for the two dimensional electron gas are much smaller than in metal.
    For Au $m^*\simeq m_0$ and the Fermi velocity is $1.4 \times 10^6$ m/s, which at 1 T gives the Larmor radius of about $R_L=8\mu$m,
    while of for a two-dimensional electron gas in GaAs $m^*=0.067m_0$ with a typical density of $n=4 \times 10^{10}/$cm$^2$, $E_F=\hbar\pi n/m^*$
    the Fermi wave vector is $0.05$ /nm and $V_F=0.086 \times 10^6$ m/, which produces $R_L=33$ nm.
\end{thebibliography}
\end{document}